\documentstyle[11pt,epsfig]{article}
\newcommand{\be}{\begin{equation}}
\newcommand{\ee}{\end{equation}}
\newcommand{\ba}{\begin{eqnarray}}
\newcommand{\ea}{\end{eqnarray}}
\newcommand{\E}{\rm E}

[1999/03/29 PSNFSS v.7.2
Times font as default roman
: S Rahtz]

\begin{document}

\title{``Slimming'' of power law tails by increasing market returns}
\thispagestyle{empty}

\author{D. Sornette$^{1,2}$ \\
$^1$ Institute of Geophysics and
Planetary Physics\\ 
and Department of Earth and Space Science\\
University of California, Los Angeles, California 90095\\
$^2$ Laboratoire de Physique de la Mati\`{e}re Condens\'{e}e\\ CNRS UMR6622 and
Universit\'{e} de Nice-Sophia Antipolis\\ B.P. 71, Parc
Valrose, 06108 Nice Cedex 2, France\\
e-mail: sornette@moho.ess.ucla.edu}

\maketitle

\vskip 1cm
\begin{abstract}

We introduce a simple generalization of rational bubble models
which removes the 
fundamental problem discovered by \cite{Luxsor} that the distribution
of returns is a power law with exponent less than $1$, in
contradiction with empirical data. The idea is that the price fluctuations
associated with bubbles must on average grow with the mean market return $r$.
When $r$ is larger than the discount rate $r_{\delta}$, the distribution of returns
of the observable price, sum of the bubble component and of the 
fundamental price, exhibits an intermediate tail with an exponent which can
be larger than $1$. This regime $r>r_{\delta}$ corresponds to a generalization
of the rational bubble model in which the fundamental price is no more
given by the discounted value of future dividends. We explain how this is possible.
Our model predicts that, the higher is the market remuneration $r$ 
above the discount rate, the larger is the power law exponent and thus
the thinner is the tail of the distribution of price returns.

\end{abstract}

\newpage

\newpage

\section{The fundamental constraint on distribution of returns of rational bubbles}

Since the publication of the original contributions on rational expectations (RE)
bubbles by \cite{Blanchard1} and \cite{Blanwat}, a huge literature
has emerged on theoretical refinements of the original concept and the empirical
detectability of RE bubbles in financial data (see 
\cite{Camerer} and \cite{adamsz}, for surveys of this literature). 
\cite{Blanwat} proposed a model with periodically collapsing bubbles in which  
the bubble component of the price follows an
exponential explosive path (the price being multiplied
by $a_t={\bar a}>1$) with probability $\pi$ and collapses to zero 
(the price being multiplied
by $a_t=0$) with probability $1 - \pi$. It is clear that, in this model,
a bubble has an exponential 
distribution of lifetimes with a finite average lifetime $\pi/(1-\pi)$. 
Bubbles are thus transient phenomena.
In order to allow for the start of new bubble after the collapse, a stochastic
zero mean normally distributed component $b_t$ is added to the systematic part of $B_t$.

Recently, \cite{Luxsor} studied the implications of the bubble models for
the unconditional distribution of prices, price
changes and returns resulting from a more general discrete-time formulation of
rational speculative bubbles:
\be
B_{t+1} = a_t B_t + b_t,    \label{eq1}
\ee
where $a_t$ can take arbitrary values 
and are i.i.d. random variables drawn from some non-degenerate
probability density function (pdf)
$P_a(a)$. The model can also be generalized by considering
non-normal realizations of $b_t$ with distribution
$P_b(b)$ with $\E[b_t]=0$. In addition, the additive term is one way among many (see 
\cite{Sorcont} for alternatives) to nucleate the bubble from a non-zero value.
Indeed, most of the time, $a_t<1$ and $B_t$ fluctuates with a scale set by $b_t$.
When an amplification phase starts up, $B_t$ increases exponentially from
this initial value and the additive contributions of succeeding $b_t$ becomes
subdominant in the exponential growth.
In (\ref{eq1}), $B_t$ denotes the difference
between the observed price and the fundamental price which defines the bubble
component.
The term ``bubble'' refers to the regimes when $B_t$ explodes exponentially
under the action of successive multiplications by factor $a_t, a_{t+1}, ...$ 
with a majority of them
larger than $1$ but different, thus adding 
an additional stochastic component to the standard model of
\cite{Blanwat}. 

Denoting $\E[.]$ the expectation operator,
provided $\E[\ln a] < 0$ (stationarity
condition) and if there is a number $\mu$ such that
$0 < \E[|b|^{\mu}] < +\infty$, such that
\be
\E[|a|^{\mu}] = 1    \label{nfakka}
\ee
and such that
$\E[|a|^{\mu} \ln |a|] < +\infty$, then the tail of the distribution of
$B$ is asymptotically (for large $B$'s) a power law \cite{Kesten,Goldie}
\be
P_B(B) ~dB \approx {C  \over |B|^{1+\mu}}~dB~,   \label{fkaka}
\ee
with an exponent $\mu$ given by the real positive solution of (\ref{nfakka}).
Rational expectations require in addition that
the bubble component in asset prices obeys
\be
\delta \E[B_{t+1}] = B_t    \label{bjqjak}
\ee
where $\delta$ is the discount factor $< 1$. This implies
\be
\E[a] = 1/\delta  >1~.
\ee
Since the function $\E[|a|^{\mu}]$ is upward convex, 
\cite{Luxsor}
showed that this automatically enforce $\mu < 1$.
It is easy to show that the distribution of price differences has the same
power law tail
with the exponent $\mu<1$ and the distribution of returns is dominated by
the same power-law over an extended range of large returns \cite{Luxsor}. 
Although power-law
tails are a pervasive feature of empirical data, these characterizations
are in strong disagreement with the usual empirical estimates which 
find $\mu$ in the range $3-5$
\cite{devries,Lux,Pagan,Guillaume1,Gopikrishnan}.
\cite{Luxsor} concluded that
exogenous rational bubbles are thus hardly reconcilable with some of the stylized
facts of financial data at a very elementary level. This result
has been extended to multi-dimensional bubbles involving an arbitrary
number of coupled assets \cite{MalSor}.

Here, we provide a simple and natural extension of model (\ref{eq1}) that removes
this constraint $\mu < 1$ and thus the discrepancy with empirical analysis. 
The key to understanding intuitively
how the result of  \cite{Luxsor} derives is to realize that the 
rational expectations condition (\ref{bjqjak}) means that the
bubble price grows locally exponentially with a instantaneous
growth rate equal to the discount rate
\be
r_{\delta} = - \ln \delta > 0~.     \label{fnnflq}
\ee
Since the stochastic auto-regressive equation (\ref{eq1}) describes
a stationary process when the condition $\E[\ln a] < 0$ is satisfied,
the only possibility to reconcile the non-stationary exponential growth with
the stationary regression is that the distribution has no mean, which indeed
occurs only for $\mu < 1$. 
In practice for any finite time intervals, the absence of a mean
implies that averages grow with the size of the time window.
This remark points to the remedy that we now present.

\section{Generalization and breakdown of the constraint}

\subsection{Rational expectation bubble model}

For the following discussion, it
is useful to recall that pricing of an asset under
rational expectations theory is based on the two following hypothesis: 
the rationnality of the agents and the ``no-free lunch'' condition.

Under the rationnal expectation condition, the best estimation of the
price $p_{t+1}$ of an asset at time $t+1$ viewed from time $t$ is
given by the expection of $p_{t+1}$ conditionned upon the knowledge of
the filtration $\{ {\mathcal F}_t \}$ (i.e. sum of all available information
accumulated) up to time $t$ : $E[p_{t+1}| {\mathcal F}_t]$.

The ``no-free lunch'' condition imposes that the expected returns of every assets
are all equal under a given probability measure $Q$ equivalent
to the historical probability measure $P$. In particular,
the expected return of each asset is equal to the return $r_{\delta}$ of the
risk-free asset (which is assumed to exist), and thus the probability measure $Q$
is  named the {\it  risk neutral probability measure}.

Puting together these two conditions, one is led to the following
valuation formula for the price $p_t$~:
\be
\label{eqfundprice}
p_t = \delta \cdot \E_{Q}[p_{t+1}|{F}_t] + d_t ~ ~~~~~~~
\forall \{p_t\}_{t \geq 0}~,
\ee
where $d_t$ is an exogeneous ``dividend'', 
and $\delta = (1+r_{\delta})^{-1}$ is the discount factor.
The first term in the r.h.s. quantifies the usual fact
that something tomorrow is less valuable than today by a factor called the
discount factor. Intuitively,
the second term, the dividend, is added to express the fact that
the expected price tomorrow has to be decreased by the dividend since the
value before giving the dividend incorporates it in the pricing.

The ``forward'' solution of (\ref{eqfundprice}) is well-known to be the fundamental price
\be
\label{eqnsolfund}
p_t^f = \sum_{i=0}^{+\infty} \delta^i \cdot \E_{Q}[d_{t+i}|{F}_t]~.
\ee
It is straightforward to check by replacement that the sum of the 
forward solution (\ref{eqnsolfund})
and of an arbitrary component $B_t$
\be
p_t  = p_t^f +  B_t~,
\label{jgjala}
\ee
where $B_t$ has to obey the single condition of being an arbitrary martingale:
\be
B_t = \delta \cdot \E_{Q}[B_{t+1}|{F}_t]    
\label{aaajqjak}
\ee
is also the a solution of (\ref{eqfundprice}). In fact, it can be
shown \cite{GouLaMon}  that (\ref{jgjala}) is the general
solution of (\ref{eqfundprice}).

Here, it is important to note that, in the framework of the Blanchard
and Watson model, the speculative
bubbles appear as a natural consequence of the valuation formula
(\ref{eqfundprice}), i.e., of the no free-lunch condition and the 
rationnality of the agents.
Thus, the concept of bubbles is not an addition to the theory, as
sometimes believed, but is entirely embedded in it.

\subsection{Model and predictions}

Consider again $B_t$ as the price difference between
observed price and the fundamental price.
We propose the following extension of the model initially defined by (\ref{eq1})
which consists in introducing an exponentially growing additive term
$b_t \to e^{rt} b_t$ with $r > 0$
such that the dynamics of $B_t$ is
\be
B_{t+1} = a_t B_t + e^{rt} b_t,   ~. \label{eq2}
\ee
The justification for modifying the additive term into $e^{rt} b_t$ is
the following. As we have seen, the additive term represents the background of
``normal'' fluctuations around the fundamental price. As we recalled above in the
definition of the standard RE model,
the additive term ensures that bubbles can nucleate spontaneously. If the economy
and the market exhibit a long-term growth rate $r$, their 
``normal'' fluctuations have also to grow with the same growth rate
in order to remain stationary in relative value. The factor $e^{rt}$ in the
additive term $e^{rt} b_t$ thus reflect nothing but the average exponential
growth of the underlying economy. In this framework, for the model to make sense,
we need to impose an additional condition $\E[\ln a] < r$ that the average
growth rate $\E[\ln a]$ of $B_t$ is smaller than the growth rate $r$
of the background economy. This condition ensures that most of time $B_t$
is a ``normal'' fluctuation with stationary 
average relative deviations from the fundamental price. Only intermittent
exponential amplifications will create transient bubbles that will
be significantly above this (exponentially growing) background.

We impose again $\E[b_t] = 0$ and $b_t$ is a white noise process.
The rational expectations (\ref{bjqjak})
leads, as before \cite{Luxsor}, to
\be
\E[a] = \delta^{-1}~.   \label{nchjakak}
\ee
We introduce the reduced price variable $A_t$ such that 
\be
B_t = e^{rt}~A_t~.  \label{cdjfalkas}
\ee
Equation (\ref{eq2}) then reads
\be
A_{t+1} = a_t e^{-r} A_t + e^{-r} b_t~,   \label{bgkaaA}
\ee
which is of the standard form (\ref{eq1}) with stationary coefficients
and obeys the usual conditions. Note that the rational expectations
condition (\ref{nchjakak}) translates into 
$\E[a_t e^{-r}] =  \delta^{-1} e^{-r}$ which can now be smaller than $1$
(see below).

Note that the condition $\E[\ln a] < r$ ensures that
$\E[\ln (a e^{-r})] < 0$ which is now the stationarity condition for the process
$A_t$ defined by (\ref{bgkaaA}). The conditions
$0 < \E[|e^r b_t|^{\mu}] < +\infty$ (which is the same condition
$0 < \E[|b_t|^{\mu}] < +\infty$ as before) and
the solution of 
\be
\E[|a e^{-r}|^{\mu}] = 1    \label{mnancac}
\ee
together with the constraint
$\E[|a e^{-r}|^{\mu} \ln |a e^{-r}|] < +\infty$ (which the same
as $\E[|a|^{\mu} \ln |a|] < +\infty$) leads to an asymptotic
power law distribution for the reduced price
variable $A_t$ of the form $P_A(A) \approx C_A/|A|^{1+\mu}$, where
$\mu$ is the real positive solution of (\ref{mnancac}).
Note that the condition $\E[\ln (a e^{-r})] < 0$ which is
$\E[\ln (a)] < r$ now allows for positive average growth rate of the product
$a_t a_{t-1} a_{t-2} ... a_2 a_1 a_0$.

\cite{Luxsor} have demonstrated rigorously
that, if $A_t$ is distributed with a power law distribution with
exponent $\mu$, then its increments $A_t-A_{t-1}$ are also 
distributed according to a power law with the same exponent $\mu$.
Here, we are interested in the distributions of the returns of the price
which we now discuss.

\subsection{Distribution of returns}

In order for the model to be meaningful with respect to returns,
we need to recognize that the observable market price is  
the sum of the bubble component $B_t$ and of a ``fundamental''
price $p_t^f$
\be
p_t = p_t^f + B_t~.   \label{jhgnakaa}
\ee
Then,
\be
p_{t+1} = p_{t+1}^f + a_t B_t + e^{rt} b_t = a_t p_t + 
(e^r - a_t) p_t^f + e^{rt} b_t~,  \label{jfkala}
\ee
where we have assumed that the economy
and the market exhibit a long-term growth rate $r$, i.e.,  
$p_{t+1}=e^r p_{t}$, leading to
\be
p_t^f = p_0 e^{rt}~.  \label{hflklx}
\ee

Expression (\ref{jfkala}) allows us to make concrete the meaning of
the additive term $e^{rt} b_t$ previously introduced in (\ref{eq2}).
Indeed, we pointed out that the additive term represents the background of
``normal'' fluctuations around the fundamental price (\ref{hflklx}). 
Their ``normal'' fluctuations have also to grow with the same growth rate
in order to remain stationary in relative value. This is shown by
replacing $p_t^f$ in (\ref{jhgnakaa}) by $p_0 e^{rt}$ given in 
(\ref{hflklx}), which leads to
\be
p_t = e^{rt} \left(p_0 + A_t \right)~.
\ee
In addition, replacing $p_t^f$ in (\ref{jfkala}) again by $p_0 e^{rt}$ leads to
\be
p_{t+1} = a_t p_t + e^{rt} [ p_0 (e^r - a_t) +  b_t ]~.  \label{jfkaaaa}
\ee
The expression (\ref{jfkaaaa}) has the same form as (\ref{eq2}) with
a different additive term $[p_0 (e^r - a_t) +  b_t ]$ replacing $b_t$.
The structure of this new additive term makes clear the origin of the
factor $e^{rt}$ introduced in (\ref{eq2}): as we said, it reflects 
nothing but the average exponential
growth of the underlying economy. The contributions $e^{rt}  b_t$ are then nothing
but the fluctuations around this average growth.

This novel additive term brings in a new feature which turns out to be essential
to describe realistic return time series:
it has a non-fluctuating component which ensures
that $p_t$ grows on the average exponentially
with time. For positive bubbles $A_t>0$ or for large fundamental price
$p_0>|A_t|$, a return $\ln(p_{t+1}/p_{t})$ 
can not be made artificially
large by the approach of $p_{t}$ arbitrarily close to $0$.
In constrast, the pure bubble component $B_t$ 
can cross $0$ an arbitrary number of times, as shown in figure 1. Close to 
these crossings, the returns $[B_{t+1}-B_t]/B_t$ of the bubble
component are very large since the denominator is close to $0$.
Since such approach of $B_t$ to $0$
occurs with a uniform probability in the vicinity of $0$, the
distribution of $1/B_t$ is determined by the rule of change of probability
density under a change of variable and is found to be
a power law with exponent equal to $1$: this
is one of the possible mechanisms known to generate
power law distributions known as the ``power law change of variable close to the origin''
\cite{mybook} (see Chap.~14). In the context of price returns, this is an 
artifact as the observable price contains the additional contribution
of the fundamental price and therefore cannot go close to zero.

The observable return is
\be
R_t = {p_{t+1} - p_t \over p_t} = {p_{t+1}^f - p_t^f + 
B_{t+1} - B_t \over p_t^f + B_t} = \chi_t 
\left({p_{t+1}^f - p_t^f \over p_t^f} + {B_{t+1} - B_t \over p_t^f}\right)
= \chi_t \left( r + {A_{t+1} - A_t \over p_0} \right)~,
 \label{jfalala}
\ee
where 
\be
\chi_t = {p_t^f \over p_t^f + B_t} = {1 \over 1 + (A_t/p_0)}~.
\label{jfoqloaq}
\ee
In order to derive the last equality in the right-hand-side of (\ref{jfalala}),
we have used the definition of the return of the fundamental price (neglecting
the small second order difference between $e^r-1$ and $r$) and we have
combined (\ref{cdjfalkas}) and (\ref{hflklx}).
Expression (\ref{jfalala}) shows that the distribution of returns $R_t$
of the observable prices is the same as that of the product of the random
variable $\chi_t$ by $r + (A_{t+1} - A_t)/p_0$. Now, the tail of the
distribution of $r + (A_{t+1} - A_t)/p_0$ is the same as the tail of the distribution
of $A_{t+1} - A_t$, which is a power law with exponent $\mu$ solution 
of (\ref{mnancac}), as shown rigorously by \cite{Luxsor}.

It remains to show that the product of this variable $r + (A_{t+1} - A_t)/p_0$
by $\chi$ has the same tail behavior as $r + (A_{t+1} - A_t)/p_0$ itself.
If $r + (A_{t+1} - A_t)/p_0$ and $\chi$ were independent, this would follow
from results in \cite{Breiman} who demonstrates that for
two independent random variables $\phi$ and $\chi$ with 
Proba$(|\phi|>x)\approx c x^{-\kappa}$  and 
E$[\chi^{\kappa+\epsilon}]<\infty$ for some $\epsilon > 0$, the
random product $\phi \chi$ obeys Proba$(|\phi \chi|>x)\approx 
{\rm E}[\chi^{\kappa}] x^{-\kappa}$. 

 $r + (A_{t+1} - A_t)/p_0$ and $\chi$ are not independent as both contain
a contribution from the same term $A_t$. However, when $A_t<<p_0$, 
$\chi$ is close to $1$ and the previous result should hold. The 
impact of $A_t$ in $\chi$ becomes important when $A_t$ becomes
comparable to $p_0$.

It then convenient to rewrite
(\ref{jfalala}) using (\ref{jfoqloaq}) as
\be
R_t = {r \over 1+ (A_t/p_0)} + {A_{t+1} - A_t \over p_0 + A_t}
={r \over 1+ (A_t/p_0)} + {(a_t e^{-r} -1) A_t + e^{-r} b_t \over p_0 + A_t}~.
\label{jfjncal}
\ee
We can thus distinguish two regimes:
\begin{itemize}
\item for not too large values of the reduced bubble term $A_t$, specifically
for $A_t < p_0$, the denominator $p_0+ A_t$
changes more slowly than the numerator of the second term, 
so that the distribution of returns will
be dominated by the variations of this numerator 
$(a_t e^{-r} -1) A_t + e^{-r} b_t$ and, hence, will follow
approximately the same power-law as for $A_t$, according to the results of
\cite{Breiman}.

\item For large bubbles, $A_t$ of the order of or greater than $p_0$, 
the situation changes, however: from (\ref{jfjncal}), we
see that when the reduced bubble term $A_t$ increases without bound, the first
term $r/(1+ (A_t/p_0))$ goes to $0$ while the second term becomes asymptotically
$a_t e^{-r} -1$. This leads to the existence of an absolute upper bound for the
absolute value of the returns given by
\be
{\rm max}|R_t| = 
{\rm max}\{|{\rm min}(a_t e^{-r} - 1)|, |{\rm max}(a_t e^{-r} - 1)|\}~.					
\label{jywujjs}
\ee
\end{itemize}
To summarize, we expect 
that the distribution of returns will therefore follow a power-law with the
same exponent $\mu$ as for $A_t$, but with a finite cut-off given in 
equation (\ref{jywujjs}).

Consider
the illustrative case where the multiplicative factors $a_t$ are
distributed according to a log-normal distribution such that
$\E[\ln a] = \ln a_0$ (where $a_0$ is thus the most probable value taken
by $a_t$) and of variance $\sigma^2$. Then, 
\be
\E[|a e^{-r}|^{\mu}] = \exp\left[ -r\mu + \mu \ln a_0 + \mu^2 {\sigma^2 \over 2}\right]~.
\label{gfjaa}
\ee
Equating (\ref{gfjaa}) to $1$ to get $\mu$ according to equation (\ref{mnancac}) gives 
\be
\mu = 2 {r - \ln a_0 \over \sigma^2} = 
{r - \ln a_0 \over r_{\delta} - \ln a_0}  = 1 + {r - r_{\delta}
\over r_{\delta} - \ln a_0} ~.   \label{fjbalal}
\ee
We have used the notation (\ref{fnnflq}) for the discount rate defined
in terms of the discount factor.
The second equality in (\ref{fjbalal}) results from (\ref{nchjakak}) using 
$\E[a] = a_0 e^{\sigma^2/2}$.

First, we retrieve the result \cite{Luxsor} that $\mu < 1$ for the initial RE model 
(\ref{eq1}) for which $r=0$ and $\ln a_0 < 0$. 
However, as soon as $r > r_{\delta} = -\ln \delta$, we get 
\be
\mu >1 ~,   \label{jtlwlal}
\ee
and $\mu$ can take arbitrary values. 
Technically, this results fundamentally from the structure of (\ref{eq2}) in which
the additive noise grows exponentially
to mimick the growth of the bubble which alleviates the bound
$\mu<1$. Note that $r$ does not need to be large
for the result (\ref{jtlwlal}) to hold. Take for instance 
an annualized discount rate $r^y_{\delta}=2\%$, an
annualized return $r^y=4\%$ and $a_0=1.01$.
Expression (\ref{fjbalal}) predicts $\mu = 3$, which is compatible with
empirical data.

\subsection{Implications of the condition $r > r_{\delta}$} 

When the price fluctuations
associated with bubbles grow on average with the mean market return
$r$, we find that the exponent of the power
law tail of the returns is no more bounded by $1$ as soon as $r$
is larger than the discount rate $r_{\delta}$ and can take essentially
arbitrary values. As can be seen from equation (\ref{eqnsolfund}),
this condition
$r > r_{\delta}$ corresponds to the unsolved regime
in fundamental valuation theory where the forward valuation solution
(\ref{eqnsolfund}) loses its meaning, as discussed recently in \cite{Sponsym}. 
Indeed, changing $d_{t+i}$ into $d_{t+i} e^{r i}$ to account for
the growth of dividend associated with the growth of the fundamental price, 
the sum in (\ref{eqnsolfund}) then behaves as 
$\sum_{i=0}^{+\infty} [\delta e^{r}]^i$ which diverges for $\delta e^{r} \geq 1$, i.e.,
$r > r_{\delta}$ (neglecting the difference between $e^r$ and $1+r$). 

Two attitudes can be taken with respect to
the condition $r > r_{\delta}$, entailing two
different modeling strategies.

\subsubsection{The ``hard-line'' attitude}

The ``hard-line' attitude (summarized from private exchanges with
T. Lux) can be described as follows.
The assumption $r > r_{\delta}$ corresponds to abandon
the background of the rational bubble theory which is the rational 
valuation formula which fails. Once this starting point is abandoned, there is
no theory of rational bubbles anymore as the transition from 
(\ref{eqnsolfund}) to (\ref{jgjala}) with (\ref{aaajqjak}) is
no more defined. The question of 
whether the results of \cite{Luxsor} are saved or changed then becomes 
irrelevant. In other words, giving up the fundamental valuation formula,
the argument goes,
would mean that we give up ``rationality'' (in the very 
limited and special sense of the rational valuation formula and 
similar theories in economics). We would then find ourselves within the 
realm of not-fully-rational behavior which implies that we have 
already given up also the possibility of having ``rational 
speculative bubbles''. This does not preclude any kind of near-rational 
bubbles like those in the agents-based models  (see for instance
\cite{LuxMarchesi,Staubook,Iori,Solomon2,sorstautak}) but
the strictly rational bubbles cease to exist.
To sum up the hard-line attitude, the theory of RE bubbles is of no interest 
anymore when one gives up the hypothesis of rational valuation because 
there is then readily available a universe of possible alternative 
bubble theories.

\subsubsection{The ``intrinsic RE bubble'' model}

The other attitude advocated here and elsewhere\\
\cite{JSL,JLS,Sormal,hyperbubblejorgen} decouples the rational
valuation formula from the RE bubble model, which we term
the ``intrinsic RE bubble'' model.
It is true that, in the
context of rational pricing based on the flow of dividend with infinite time
horizon, an exponential growth of the price is associated with the same
exponential growth of the dividend. This leads to an inconsistency as the 
valuation formula leads to a diverging price if the
growth rate is larger than the discount rate. As discussed in \cite{Sponsym},
the price is of course not infinite, only the rational valuation formula
is of no use to determine it. The price
has to be defined by processes other than the dividend
valuation formula. 
\begin{itemize}

\item From analogies with statistical physics and with
the theory of bifurcations and their normal forms, in which
similar situations occurs, I proposed in \cite{Sponsym} a scenario
according to which this regime
may be associated with a spontaneous symmetry breaking phase corresponding
to a spontaneous valuation in absence of dividends. 

\item Another approach
is to realize that the fundamental valuation formula is a statement
of equilibrium. It is thus natural to interpret the 
regime $r > r_{\delta}$ where the fundamental valuation formula fails 
as intrinsically dynamical: the fundamental price becomes 
determined by dynamical processes
other than just the flow of dividends. 

\item It is possible to develop models of general equilibrium \cite{blackgeneral}
in which one can treat the growth of the economy reflected in the stock market
as the accumulation of composite capital, plus a change of available technology
that affects both the old and new capital. Capital's market value grows as people
add new units of capital, or as existing units grow in value as they become more
effective. If $K_t$ is the composite capital or market value at time $t$, $r_t$
is the return and $c_t$ is consumption, then 
$K_{t+1} = r_t K_t -c_t$. In the long run, consumption seems to grow at the same
rate as wealth, income and other measures of available resources. Thus $c_t$ is
roughly proportional to $K_t$, i.e., $c_t = \gamma_t K_t$ with $\gamma_t < r_t$
as people in general consume less that the expected return on composite capital,
leading to expected growth. In this framework \cite{blackgeneral}, dividends
are seen as depending on current and past prices rather than prices as depending
on expected future dividends as in the standard valuation formula. It thus
is perfectly possible to model a market growing at an average rate $r$
equal to the average of $r_t-\gamma_t$
which can be larger than the risk-free rate or discounting rate or inflation rate.
This does not invalidate the assumption of rational expectations, which simply
states that people make economic decisions in a way that tends to take into account
all available information bearing significantly on the future consequences of their
decisions. We can still form rational expectations from an intelligent appraisal
of circumstances, though the process behind such circumstances may be hard to discern.
An economic system doing the most
efficient possible job of reading the information being reflected in price
signals will still experience some irreducible business cycle swings and 
bubbles. This results from the fact that
the economic process contains inherent mechanisms that convert random
shocks on prices into a more persistent, short-term misreading of changing profit
opportunities, in other words the rational interpretation of noise 
can stimulate a
cumulative swing in output that will continue until the misreading is realized
and retrenchment sets in. Random shocks to prices and markets are always with us
and rational bubbles are thus naturally occurring market phases.
Burke \cite{Burke} has also established, by
dropping continuity and weakening utility representation, that commodity prices 
and consumptions can approach approximate equilibrium to
within any practical tolerance after dropping the standard general-equilibrium 
assumption that preference orders discount future consumption faster than the economy
grows.

\item Another fix is based on the 
understanding that the divergence of the
sum in (\ref{eqnsolfund}) stems from the fact that the investor
assumes an infinite time horizon; in practice, we would discount
future cash flows only up to a maximum time, say ten or thirty years at most.
As a consequence, the rational valuation formula has to be truncated
to a finite number of terms and provides a finite price.

\end{itemize}

Therefore, the apparent contradiction is resolved by
abandoning the pricing based on dividends discounted over an
infinite time horizon and replacing it by an exogeneous dynamical process.
In other words, we assume here that the fundamental price is growing 
exponentially, without reference to dividends. 
We thus propose not to ``throw away the baby with the bath'', i.e., to
decouple the fundamental pricing mechanism from the RE bubbles.
Indeed, investors use various methods for
estimating the fundamental price and the rational valuation formula
is only one among an arsenal of techniques that,
in real life situations, are usually combined.

The point of view advocated here is that it is reasonable to keep the 
RE bubble model (\ref{aaajqjak}), independently of what determines
the specific underlying fundamental price. This is actually the point of
view initially introduced by \cite{Blanchard1} and 
further developed in \cite{JSL,JLS,hyperbubblejorgen}, where the 
bubble price may follow essentially arbitrary trajectories as long as
there is a jump process (crash) with a hazard rate ensuring the 
no-arbitrage condition.

In a way, the choice between the ``hard-line'' attitude and
the ``intrinsic RE bubble'' model is 
almost a matter of philosophy with respect to the meaning of models.
Ultimately, a model is useful if it helps thinking about the problem
and if it allows one to formulate new questions. It seems interesting
to use the modelling strategy in which the RE condition is relaxed
partially by abandoning only the rational valuation formula.
Following this view point, we have the following logical sequence:
\begin{enumerate}
\item RE bubbles give an exponent $\mu>1$
for the distribution of returns only when the fundamental price 
is growing faster than the discount rate;
\item when the fundamental price 
is growing faster than the discount rate, the rational valuation formula
breaks down;
\item empirical studies show that $\mu>1$ in 
real financial time series;
\item as a consequence, since the distribution of returns
are usually constructed over rather long time histories, it
is sufficient that the fundamental valuation departs from the dividend formula at
some times to explain item 3.
\end{enumerate}
This provides a parsimonious model for the empirical distribution of returns.

\subsection{Numerical simulations}

We now present numerical simulations that illustrate these results.
We do not pretend to capture reality acurately but show nevertheless
that the stylized facts of empirical data are recovered by this simple model.
In order to make a precise comparison with empirical data, one should 
specify the possible shapes of the distributions $P_a(a)$ and $P_b(b)$ 
and one should also add the fundamental price to the bubble price $B_t$
since only the sum is observable.
Tests of the model would thus be a joigned test of the relevance of the
RE model together with an assumption on the dynamics of the fundamental price.

\begin{figure}
\begin{center}  
\epsfig{file=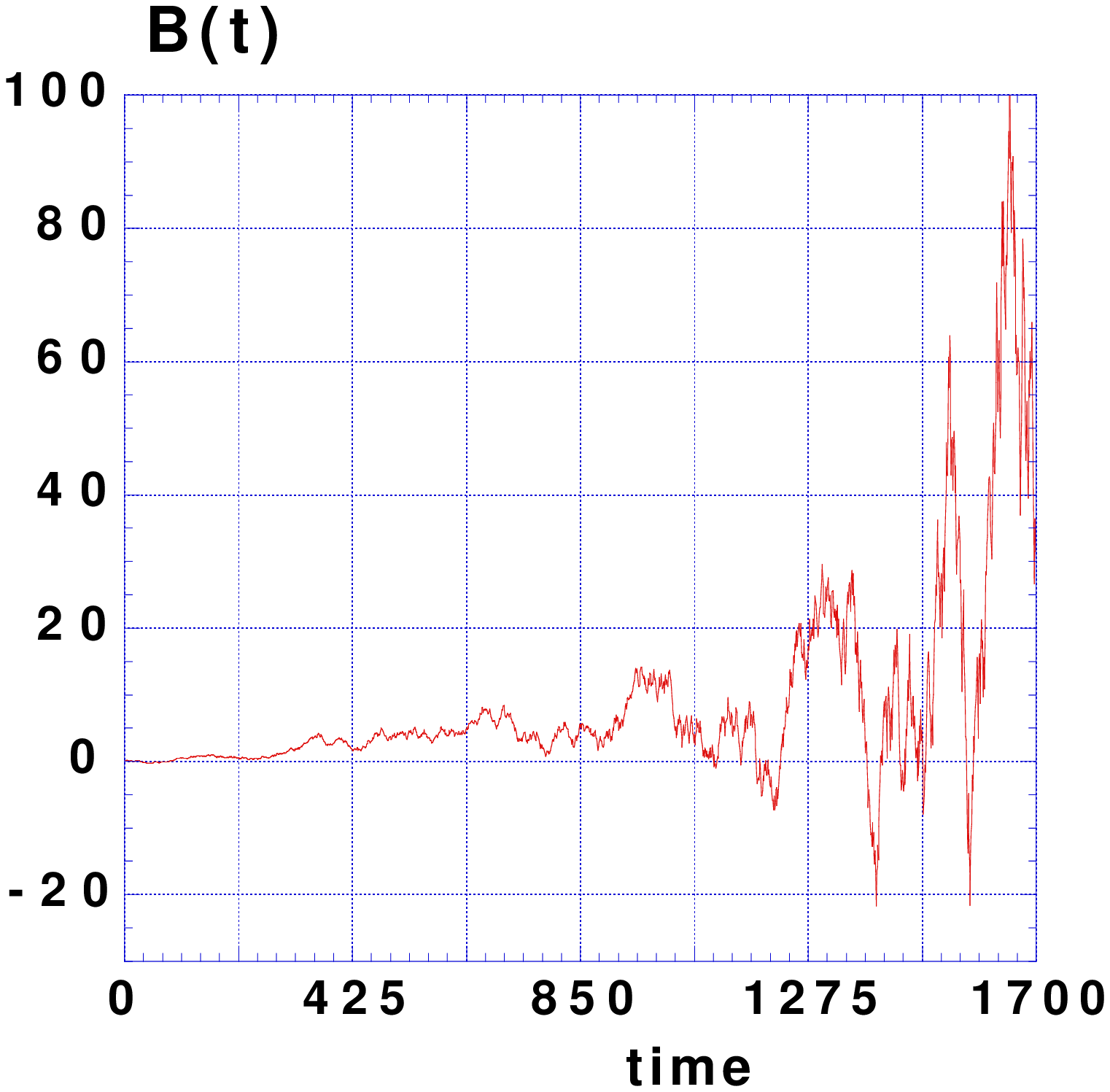,height=15cm,width=16cm}
\caption{\label{fig1}  Bubble price $B_t$ generated by 
equation (\ref{eq2}) with $r=0.33\%$. The
multiplicative factors $a_t$ are generated from the formula (\ref{jgnalna}).
$a_t$ are thus log-normal random variables with 
$\E[\ln a]=\ln a_0=0.1\%$ and
$\E[a]=1.0017$. The additive term $b_t$ is 
uniformely distributed in the interval $[-0.05; +0.05]$.
With these parameters, one time step can be interpreted as
approximately one month. The time interval shown covers
thus approximately $80$ model-years.
}
\end{center}
\end{figure}

Figure \ref{fig1} shows a typical trajectory of the price $B_t$
of a bubble generated by 
equation (\ref{eq2}) with $r=0.33\%$. This choice is such that
one time step can be approximately interpreted as one month, since the
compounded return over $12$ time steps gives a realistic yearly return
of $4\%$ ($1.0033^{12}=1.04$), as can be seen in figure 2.

The multiplicative factors $a_t$ are generated from the formula
\be
a_t=a_0 \exp[\sigma \eta_t]~,~~~~{\rm where}~~a_0=1.001~~~{\rm and}
~~\sigma = 0.0374~,  \label{jgnalna}
\ee
and the
$\eta_t$'s are centered Gaussian random variables with unit variance.
With this parameterization, $a_t$ are log-normal random variables
of variance $\sigma^2$
with ${\rm E}[\ln a]=\ln a_0=0.001$ and ${\rm E}[a]=a_0 e^{\sigma^2/2}
=1/\delta=1.0017$. Therefore, $\delta=0.9983$ and $r_{\delta}$ defined by 
(\ref{fnnflq}) is $r_{\delta}=0.17\%$.
With the values $r=0.33\%, r_{\delta}=0.17\%$ and $\ln a_0=0.1\%$, 
expression (\ref{fjbalal}) predicts an exponent $\mu = 3.3$.

The additive term $b_t$ is taken
uniformely distributed in the interval $[-0.05; +0.05]$.
Note that $b_t$ sets the scale of $B(t)$. 
The bubble price $B_t$ stays on average at $0$, as seen from (\ref{eq2})
which predicts $\E[B_t] = 0$. However, like a correlated random walk with
fat tails, it wanders around, with an appearance quite reminiscent of
active and volatile markets. 

\begin{figure}
\begin{center}  
\epsfig{file=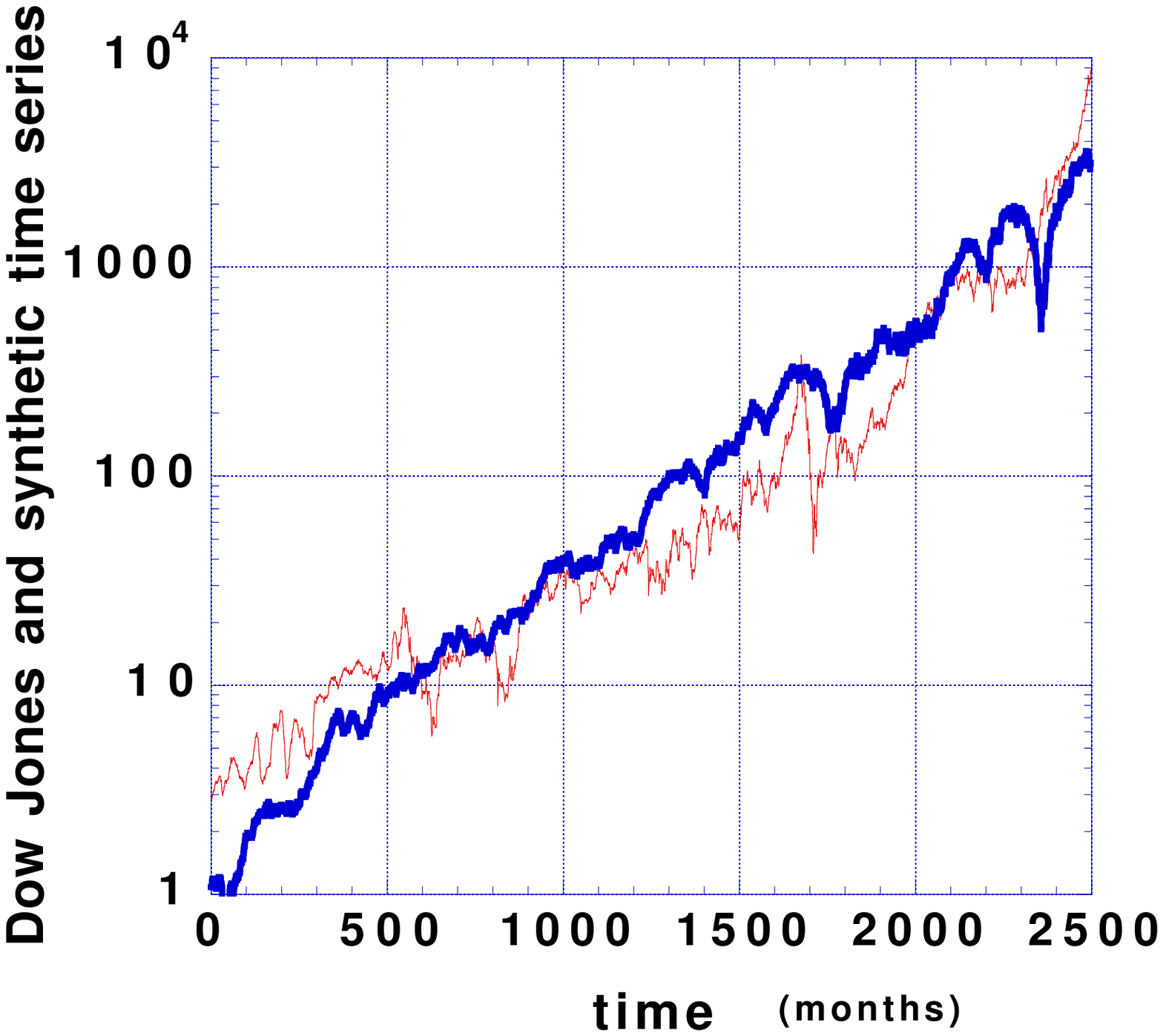,height=15cm,width=16cm}
\caption{\label{fig2} The thick line shows
the time dynamics of $B_t+p_t^f$ (bubble $+$ 
fundamental price), where we have assumed that
the fundamental price $p_t^f=e^{r t}$ follows a deterministic growth at the ``annualized''
rate of $4\%$ corresponding to the value $r=0.33\%$. 
The trajectory of the Dow
Jones Industrial Average (DJIA) extrapolated back from 1790 till 2000 is 
also shown as the thin line. 
}
\end{center}
\end{figure}

As discussed in the previous section, in order to make a meaningful
comparison with empirical data, it is necessary to add a fundamental
component to the bubble to form an observable price. To remain
as simple as possible, we follow the specification 
(\ref{jhgnakaa}) with (\ref{hflklx}), with $p_0=1$. Figure
\ref{fig2} shows the time dynamics of $B_t+e^{r t}$, where $B_t$ is the same
as shown in figure \ref{fig1} and we have assumed that
the fundamental price follows a deterministic growth at the ``annualized''
rate of $4\%$ (corresponding to the monthly
rate $r=0.33\%$) according to (\ref{hflklx}). The trajectory of the Dow
Jones Industrial Average (DJIA) extrapolated back from 1790 till sept. 2000 is 
also shown as the thin line. The Dow Jones
index was constructed by The Foundation for the Study of Cycles
($http://www.cycles.org/cycles.htm$). It is striking to observe
how the simple RE bubble model added to a simple exponentially growing
fundamental price can capture the large scale variability of the DJIA.

\begin{figure}
\begin{center}  
\epsfig{file=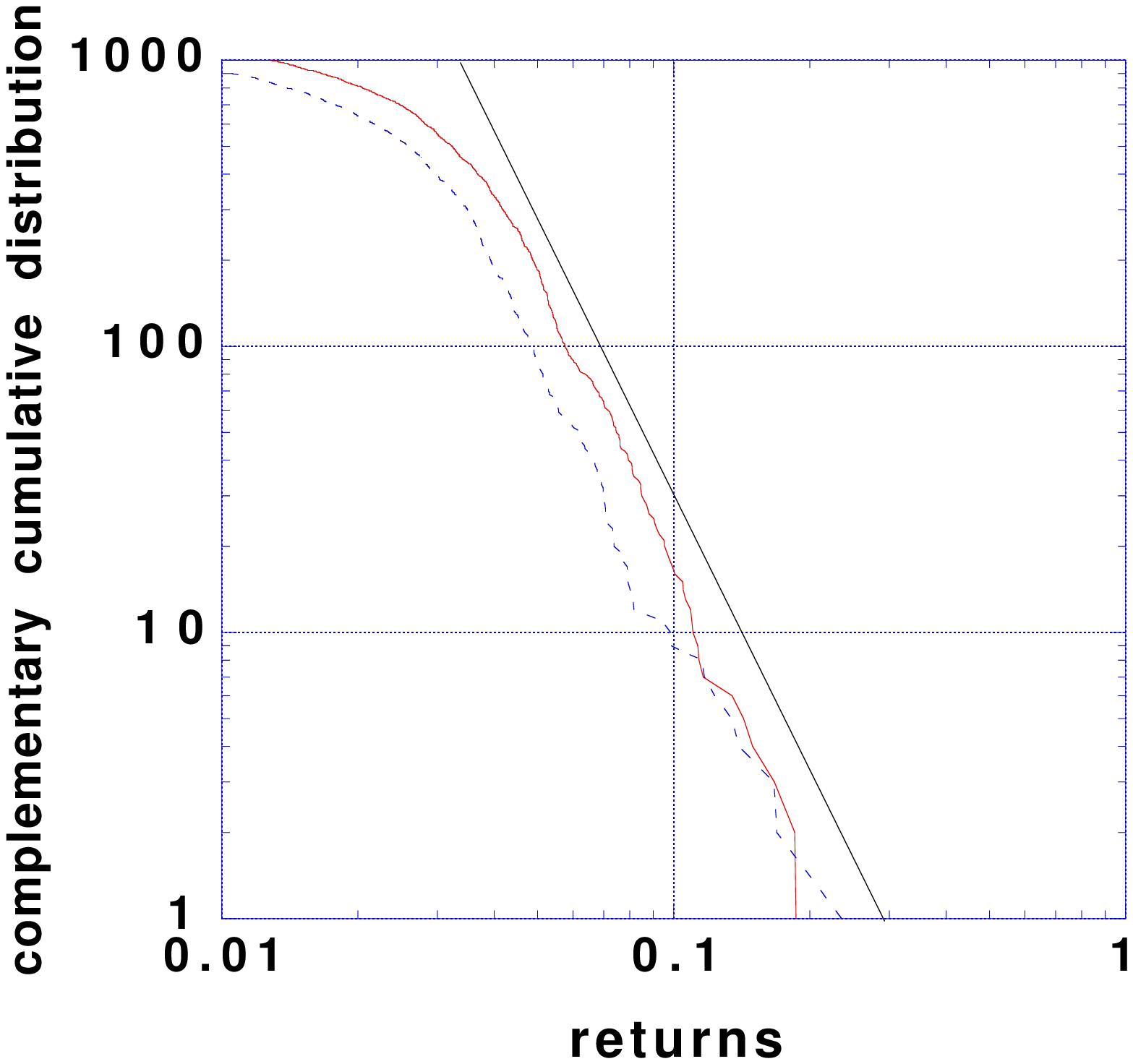,height=15cm,width=16cm}
\caption{\label{fig3} Double logarithmic scale representation of the
complementary cumulative distribution of the 
``monthly'' returns $R_t$ defined in (\ref{jfalala}) of the synthetic total
price shown in figures \ref{fig2}. 
The continuous (resp. dashed) line corresponds to the positive (resp. negative) returns.
The distribution is 
well-described by an asymptotic power law with an exponent in agreement
with the prediction
$\mu \approx 3.3$ given by the equations (\ref{mnancac})
and (\ref{fjbalal}) and shown as the straight
line. The small differences between the predicted slope and the numerically generated
ones are within the error bar of $\pm 0.3$ obtained from a standard maximum
likelihood Hill estimation.
}
\end{center}
\end{figure}

Figure \ref{fig3}
shows a double logarithmic scale representation of the
complementary cumulative distribution of the 
``monthly'' returns $R_t$ defined in (\ref{jfalala}), constructed
from the synthetic time series shown in figure \ref{fig2}.
As predicted, the distribution is 
well-described by the asymptotic power law with an exponent
in agreement with the prediction
$\mu \approx 3.3$ given by the equations (\ref{mnancac})
and (\ref{fjbalal}) and shown as the straight
line. Note that the expected cut-off at the maximum return is not observed
as the returns have not yet explored the large values of the order of $1$
in the finite-time series.

\begin{figure}
\begin{center}  
\epsfig{file=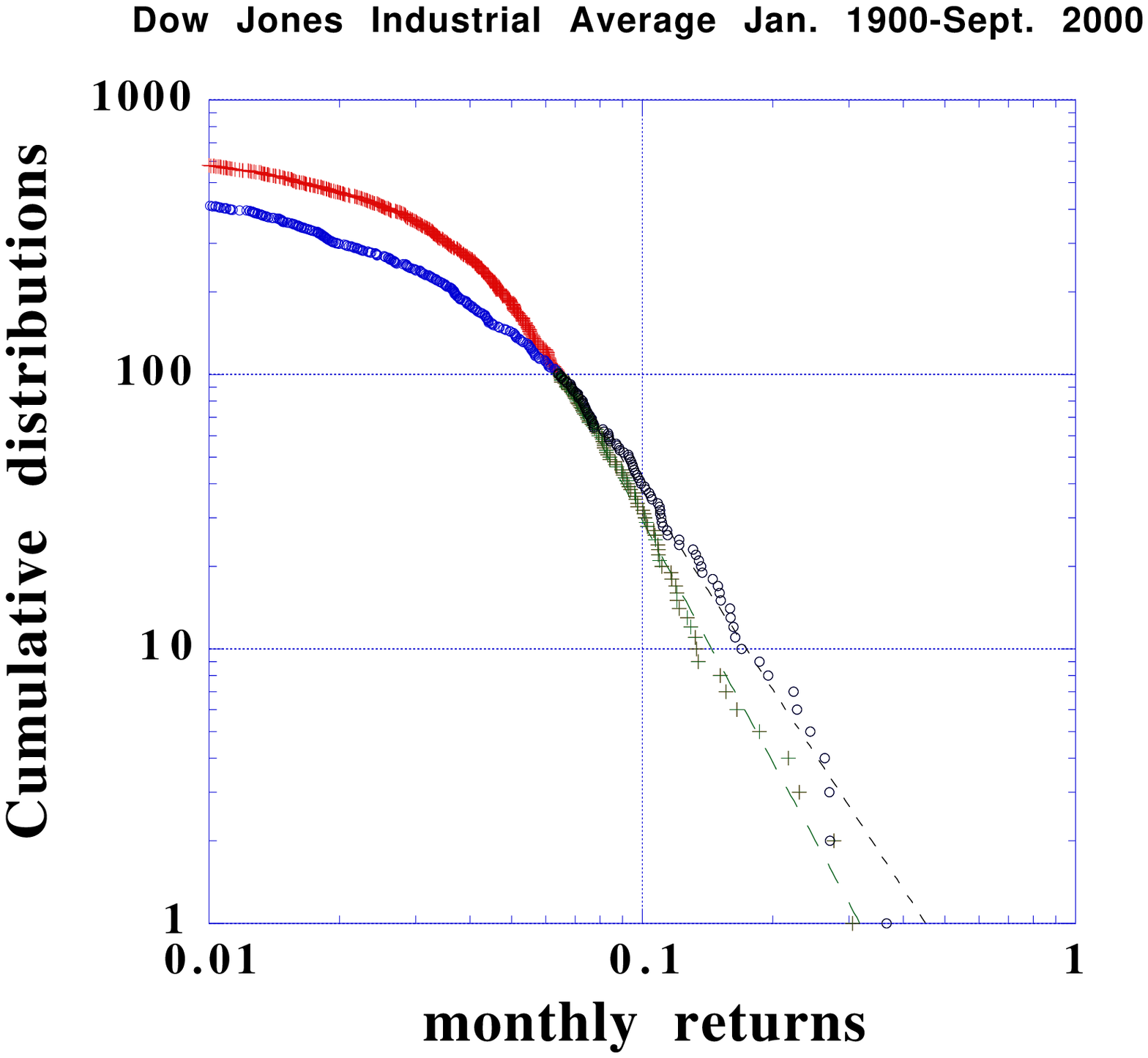,height=15cm,width=16cm}
\caption{\label{fig4} Double logarithmic scale representation of the
complementary cumulative distribution of the 
monthly returns $\ln p_{t+1}/p_t$ of Dow Jones Industrial Average price $p_t$
from Jan. 1900 till Sept. 2000. The $+$'s (resp empty circles) correspond
to positive (resp. negative) returns. The two dashes straight lines 
give the best fits to a power law for the largest $100$ positive
and negative monthly returns. For positive (resp. negative) returns, we find an
exponent $\mu_+ = 2.9 \pm 0.3$ (resp. $\mu_- = 2.4 \pm 0.3$). This is
in agreement with previous empirical
works \cite{devries,Lux,Pagan,Guillaume1,Gopikrishnan}.
}
\end{center}
\end{figure}

For comparison, figure \ref{fig4} shows the distribution of positive and negative 
returns of the Dow Jones Industrial Average price 
over the 20th century. The tails are very well quantified by power laws
with exponents respectively equal to $\mu_+ = 2.9 \pm 0.3$ and $\mu_- = 2.4 \pm 0.3$.
The error bars are estimated from the theory of maximum likelihood applied to the
Hill estimator which predict $\delta \mu/\mu = 1/\sqrt{N}$, where $N$ is the number
of data in the power law tail. Using $N \approx 100$, we find $\delta \mu \approx 0.3$.
The data does not reject the hypothesis that $\mu_+ =\mu_- \approx 2.7 \pm 0.3$.
These values are compatible with those previously reported in the 
literature on smaller time scales \cite{devries,Lux,Pagan,Guillaume1,Gopikrishnan}.
The comparison between figures \ref{fig3} and \ref{fig4} suggests
that the RE bubble model (\ref{eq2}) goes a long way into capturing
the structure of real market prices.

\section{Concluding remarks}

In the numerical example given in the previous section, the stationary process
$A_t$ with zero mean and finite variance may in principle eventually reach
the value $-p_0$ with very low but not zero probability, 
at which the observable price vanishes. Such large negative
bubbles will thus produce artifacts in the distribution of returns which, as
we demonstrated
for the returns of $B_t$, are controlled by the excursions of $p_t$ close to
$0$. The general formulation of the RE bubbles
allows in principle for both positive and negative bubbles with the
sign of each new bubble depending on that of the additive stochastic term in its
inception period. However, it is well known that there are conceptual problems
with negative bubbles \cite{diba}. Therefore, we
deliberately confined ourselves to bubbles of amplitudes smaller than the 
fundamental price. Numerical tests using only positive bubbles, obtained
by imposing a strickly positive additive term $b_t$, confirm the results
of figures 2 and 3. We thus emphasize that none of our results hinges on this
slight modification of the model and both modeling strategies are equally viable:
consider only the bubbles that do not become too negative so that the 
total observable price does not become too small or construct only positive bubbles.
This conclusion has also been checked extensively by \cite{Luxsor}.

While investors enjoy getting a larger return $r$,
this comes usually at the price of increasing risks. Here, the 
situation is different because a higher market return leads to a {\it thinner}
tail of the return distribution since the exponent $\mu$, for instance 
given by (\ref{fjbalal}), increases with $r$. Hence, increasing the average
market return $r$ decreases the extreme risks.
In this context, it is interesting to bring into focus 
the long-standing paradox that the Dow Jones average index
has been argued to exhibit an anomalously large
return, averaging $6\%$ per year over the 1889-1978 period \cite{Mehra},
which cannot be explained by any reasonable risk aversion coefficient.
What our study shows is that
the excess remuneration $r-r_{\delta}$ has an unexpected ``good''  consequence
in decreasing drastically the large risks of the market by increasing the
exponent of the asymptotic power law distribution. 
Following recent ideas in the theory of complex systems 
(see for instance \cite{mybook} and references therein),
we argue that the market has self-organized such that the excess remuneration
has reached a level which compensate for the huge risks associated 
with the intermittent bubbles created by investors. 
We conjecture that
the presently observed value of the exponent $\mu$ in the range $3-5$ 
\cite{devries,Lux,Pagan,Guillaume1,Gopikrishnan}
and the anomalous returns
of the market over long period of times are the tools that the system 
has developed to tame the fat tails that bubbles tend to create.
Extending the present model to derive a dynamics on $r$ would allow 
us to understand how this organization may proceed.

Acknowledgements: I am grateful to T. Lux and Y. Malevergne for
stimulating discussions.

\pagebreak

\pagebreak

\pagebreak 

\pagebreak 

\pagebreak 

\end{document}